\documentclass[pre,preprint]{revtex4}

\usepackage{amsfonts}
\usepackage{amsmath}
\usepackage{amssymb}
\usepackage{theorem}

\newcommand{\one}{\mbox{\tt 1}\hspace{-0.057 in}\mbox{\tt l}}
\newcommand{\natuerl}{{\mathbb N}}

\newcommand{\Tr}{\mbox{\rm\small Tr\ }}

\newcommand{\balphas}{\mbox{\boldmath $\scriptstyle \alpha$}}

\newcommand{\bbetas}{\mbox{\boldmath $\scriptstyle \beta$}}

\theoremstyle{break}

\begin{document}
\title{Initial states and decoherence of histories}
\author{Artur Scherer\footnote{the corresponding author:
a.scherer@rhul.ac.uk}}
\author{Andrei N.\ Soklakov}
\address{Department of Mathematics, Royal Holloway, University of London,
Egham, Surrey, TW20 0EX, UK.}
\begin{abstract}
We study decoherence properties of arbitrarily long histories 
constructed from a fixed projective partition of a finite 
dimensional Hilbert space. We show that decoherence of such 
histories for all initial states
that are naturally induced by the projective partition  
implies decoherence for arbitrary
initial states. In addition we generalize
the simple necessary decoherence condition 
[Scherer et al., Phys.\ Lett.\ A (2004)]   
for such histories to the case of arbitrary coarse-graining. \\ 



\end{abstract}
\date{10 May, 2004}
\maketitle

In the Copenhagen interpretation of quantum mechanics
all properties of a quantum system are defined with
respect to measurements performed by an external observer
using classical measuring devices. This interpretation,
however, cannot be used in the case of closed quantum
systems, such as the Universe as a whole. In this case any
observer must be a part of the system itself. 
A self-contained description of closed quantum systems
that does not rely on either the external observer nor
on the existence of classical devices is provided by 
the decoherent histories   
approach~\cite{Griffiths1984,Omnes1988,Gell-Mann1990,Dowker1992,Gell-Mann1992}. 
This approach predicts probabilities for quantum histories, 
i.e.\ ordered sequences of quantum-mechanical 
``propositions''. Mathematically,
these propositions are represented by projectors:
the same projectors that would define a quantum
measurement in the Copenhagen approach.
In particular, an exhaustive set of mutually exclusive propositions
corresponds to a complete set of mutually
orthogonal projectors.

Due to quantum interference, one cannot always assign 
probabilities to a set of histories in a consistent way. 
For this to be possible, the set of histories must be 
decoherent. Whether the corresponding decoherence condition 
is fulfilled or not depends on the initial state, 
the unitary dynamics of the system and the propositions 
from which the histories are constructed.
In this paper we consider histories that
are constructed from a {\it fixed} exhaustive 
set of mutually exclusive propositions, 
$\{P_\mu\}$, and investigate the question of how 
the choice of the initial state affects 
decoherence of such histories.
We show that decoherence of arbitrarily
long histories for all initial states
that are induced by the projectors
$\{P_\mu\}$ via normalization implies
the decoherence for arbitrary initial states. 
It is relevant to note that, unlike the set of 
all possible states, the set $\{P_\mu\}$ is discrete 
and may contain as few as just two elements     
(for \lq\lq yes-no'' propositions).
As an additional result, we obtain a generalization
of the simple necessary decoherence condition
that was derived for fine-grained histories 
in~\cite{Scherer2004A}. The new condition is applicable to arbitrary
coarse-grainings.

The paper is organized as follows. 
After introducing our framework we 
present the mathematical content of 
our results in the form of a theorem.
We prove the theorem, infer the 
results and conclude with a short
summary.

\noindent {\bf Definition 1:}
A set of projectors $\{P_{\mu}\}$ on a Hilbert space $\cal H$ is 
called a projective {\em partition} of $\cal H$, if $\:\forall\, \mu, 
\mu'\,:\;\:P_{\mu}P_{\mu'}=\delta_{\mu\mu'}P_{\mu}\:$ and $\:\sum_{\mu}P_{\mu}
=\one_{\cal H}$. Here, $\one_{\cal H}$ denotes the unit operator. 
We call a projective partition  {\em
  fine-grained\/} if all projectors are one-dimensional, 
i.e.,  $\,\forall\,\mu\;$
$\mbox{dim}\big(\mbox{supp}(P_{\mu})\big)=1$
\footnote{The {\em support} of a Hermitian operator $A$ is defined to 
be the vector space spanned by the eigenvectors of $A$ corresponding 
to its non-zero eigenvalues.},
and {\em coarse-grained} otherwise.

\noindent{\bf Definition 2:}
Given a projective partition $\{P_{\mu}\}$ of a Hilbert space $\cal H$, 
we denote by $\mathcal{K}[\{P_{\mu}\}\,;\,k\,]:=\big\{h_{\balphas}\,:\:
h_{\balphas}=\left(P_{\alpha_{1}},P_{\alpha_{2}},
 \dots, P_{\alpha_{k}}\right)\in\{P_{\mu}\}^k\big\}$ the 
corresponding exhaustive set of mutually exclusive histories of length
$k$. Histories are thus defined to be ordered sequences of projection
operators, corresponding to quantum-mechanical propositions. 
Note that we restrict ourselves to histories constructed from a 
{\em fixed} projective partition: the projectors $P_{\alpha_{j}}$ within the 
sequences are all chosen from the same partition for all 
\lq\lq times''  $j=1,\ldots,k$.

\noindent{\bf Definition 3:}
Given a Hilbert space $\cal H$ and a projective partition
$\{P_{\mu}\}$ of $\cal H$, we denote by $\mathcal{S}$ the 
set of all density operators on $\cal H$ and by
$\mathcal{S}_{\{P_{\mu}\}}$
the discrete set of {\em \lq\lq partition states''} induced by the partition 
$\{P_{\mu}\}$ via normalization:
\begin{equation}  
\mathcal{S}_{\{P_{\mu}\}}:=
\Big\{\frac{P_{\nu}}{\Tr[P_{\nu}]}\,:\,P_{\nu}
\in\{P_{\mu}\}\Big\}\,.
\end{equation}

An initial state $\rho\in\mathcal{S}$ and a unitary  
dynamics generated by a unitary map $U:\cal H\rightarrow\cal H$ induce 
a probabilistic structure on the event algebra 
associated with $\mathcal{K}[\{P_{\mu}\}\,;\,k]$, 
if certain consistency conditions are fulfilled.
These are given in terms of properties of the {\em decoherence functional\/}
$\mathcal{D}_{U,\,\rho}\,[\cdot,\cdot]$ on 
$\mathcal{K}[\{P_{\mu}\}\,;\,k\,]\times
\mathcal{K}[\{P_{\mu}\}\,;\,k\,]\,$, defined by 
\begin{equation} 
\mathcal{D}_{U,\,\rho}\,[h_{\balphas},h_{\bbetas}]:=
\mbox{Tr}\left[C_{\balphas}\,\rho\,C_{\bbetas}^{\dagger}\right]\:,
\end{equation}
where 
\begin{eqnarray} \label{Classoperators}
C_{\balphas}& :=&\left(U^{\dagger\,k}P_{\alpha_k}U^k\right)
\left(U^{\dagger\,k-1}P_{\alpha_{k-1}}U^{k-1}\right)\dots
 \left(U^{\dagger}P_{\alpha_1}U\right)\nonumber\\
&=& U^{\dagger\,k}P_{\alpha_k}UP_{\alpha_{k-1}}U\dots
  P_{\alpha_2}UP_{\alpha_1}U\;.
\end{eqnarray}
The set $\mathcal{K}[\{P_{\mu}\}\,;\,k\,]$ is said to be 
{\em decoherent\/} or {\em consistent\/} with respect to a given 
unitary map $U:\cal H\rightarrow\cal H$ and a given initial 
state $\rho\in\mathcal{S}$, if 
\begin{equation}  \label{eq:consistency}
\mathcal{D}_{U,\,\rho}\,[h_{\balphas},h_{\bbetas}]
\propto \delta_{\balphas\bbetas}\equiv
\prod_{j=1}^k\delta_{\alpha_j \beta_j}
\end{equation}
for all $h_{\balphas},h_{\bbetas}\in\mathcal{K}[\{P_{\mu}\}\,;\,k\,]$.
These are the consistency conditions. 
If they are fulfilled, probabilities  
may be assigned to the histories   
and are given by the diagonal elements 
of the decoherence functional, 
$p[h_{\balphas}]=\mathcal{D}_{U,\,\rho}\,[h_{\balphas},h_{\balphas}]$.

What we have just described is a slightly simplified version of the 
general decoherent histories formalism. In general, both the partition 
and the unitary may depend on the parameter $j=1,\ldots,k$. Our
setting based on a fixed partition and a fixed unitary is motivated 
by the analogy with the classical symbolic 
dynamics~\cite{Scherer2004A,Alekseev1981}.
In the literature several consistency conditions of different strength 
can be found~\cite{Gell-Mann_1998}. The conditions 
given above are known as {\em medium decoherence\/}~\cite{Gell-Mann1992,Diosi2004}.

\noindent{\bf Theorem:}
{\it Let a projective partition}  $\{P_{\mu}\}$ 
{\it of a finite dimensional Hilbert space $\cal H$ and a unitary 
map} $U$ {\it on} $\cal H$ {\it be given. Then the following three 
statements are equivalent:}
\begin{eqnarray}
\label{eq:decoherence_all_P_{mu}}
&(a)&
\forall \,\rho\in
\mathcal{S}_{\{P_{\mu}\}}\;
\forall\, k\in\hspace{-0.5mm}\natuerl \;\;\forall\,h_{\balphas},
h_{\bbetas}\in\mathcal{K}[\{P_{\mu}\}\,;\,k\,]:
\;\mathcal{D}_{U,\rho}\,[h_{\balphas},h_{\bbetas}]
\propto  \delta_{\balphas\bbetas}\nonumber\\
\label{eq:commutativity_P}
&(b)&
\forall\,P_{\mu'}, P_{\mu''}\in\{P_{\mu}\}\;\,\forall\, n\in\natuerl
\,:\;\;\left[U^nP_{\mu'}(U^{\dagger})^n\,,\,P_{\mu''}\right]=0\nonumber\\
\label{eq:decoherence_all_rho}
&(c)&
\forall \,\rho\in\mathcal{S}\;\;
\forall\, k\in\hspace{-0.5mm}\natuerl \;\;\forall\,h_{\balphas},
h_{\bbetas}\in\mathcal{K}[\{P_{\mu}\}\,;\,k\,]:
\;\mathcal{D}_{U,\rho}\,[h_{\balphas},h_{\bbetas}]
\propto  \delta_{\balphas\bbetas}\;.\nonumber
\end{eqnarray}

\vspace{2mm}

\noindent{\bf Proof:}
We will prove the theorem by showing that (a) implies (b), 
(b) implies (c), and (c) implies (a). The last implication, 
(c)$\Rightarrow$(a), is trivial, and the second implication, 
(b)$\Rightarrow$(c), can be easily shown using the notation 
of Eq.~(\ref{Classoperators}). It remains to prove the 
implication (a)$\Rightarrow$(b).

The proof is constructed as follows. 
We first show that the proposition 
\begin{eqnarray} \label{eq:loopcondition}
&&\forall
  \,\rho\in\mathcal{S}_{\{P_{\mu}\}}\:\,
\forall\, n\in\natuerl\:\:
\forall\,\mu_0,\mu',\mu''\quad\mbox{with}\quad
\mu'\not=\mu'':\nonumber\\
&&\mbox{Tr}\left[P_{\mu''}(U^nP_{\mu_0}U^{\dagger\,n})P_{\mu'}
(U^n\rho U^{\dagger\,n})P_{\mu''}\right]=0\;\,
\end{eqnarray}
is a necessary consequence of the decoherence condition (a)
and then conclude that this proposition implies the commutativity condition 
(b) of the theorem.

The first part of the proof will be accomplished by contradiction, 
i.e.\ we will assume that (\ref{eq:loopcondition}) is not satisfied, 
and then show that this assumption contradicts the decoherence 
condition (a) of the theorem.

Assume condition (\ref{eq:loopcondition}) is not satisfied. 
This means there exist a partition state 
$\tilde{\rho}\in\mathcal{S}_{\{P_{\mu}\}}$, 
an integer $\tilde{n}\in\natuerl$, 
and partition-element labels $\mu_0$, $\mu'$, $\mu''$, with 
$\mu'\ne\mu''$, such that 
\begin{equation} \label{eq:Tr[...]=c} 
\mbox{Tr}\left[P_{\mu''}(U^{\tilde{n}}P_{\mu_0}
U^{\dagger\,{\tilde{n}}})P_{\mu'}
(U^{\tilde{n}}\tilde{\rho}\, U^{\dagger\,{\tilde{n}}})
P_{\mu''}\right]=c\not=0\;.
\end{equation}
This, as we will see, is in contradiction to decoherence 
condition~(a). Written out, the decoherence condition~(a) is  
\begin{equation} \label{DecCondition}
\mbox{Tr}\left[P_{\alpha_k}UP_{\alpha_{k-1}}U\dots
  P_{\alpha_1}U\,\rho_0\: U^{\dagger}P_{\beta_1}\dots
  P_{\beta_{k-1}}U^{\dagger}P_{\beta_k}\right]\propto \prod_{j=1}^k
  \delta_{\alpha_j \beta_j}
\end{equation}
for all $k\in\natuerl$, all initial states 
$\rho_0\in\mathcal{S}_{\{P_{\mu}\}}$,
and arbitrary histories $h_{\balphas}$,
$h_{\bbetas}$. Since the length $k$ of the histories is arbitrary, 
we may choose $k=q\tilde{n}$ with arbitrary $q\in\natuerl$. By summing over 
$\alpha_1,\dots,\alpha_{\tilde{n}-1},\alpha_{\tilde{n}+1},\dots,\alpha_{q\tilde{n}-1}$ and 
$\beta_1,\dots,\beta_{\tilde{n}-1},\beta_{\tilde{n}+1},\dots,\beta_{q\tilde{n}-1}$, 
and using $\sum_{\mu}P_{\mu}=\one_{\cal H}$, 
we obtain
\begin{equation}\label{nec_con_(q-1)}
\mbox{Tr}\left[P_{\alpha_{q\tilde{n}}}(U^{q-1})^{\tilde{n}} 
P_{\alpha_{\tilde{n}}}U^{\tilde{n}}\,\rho_0\: 
U^{\dagger\,\tilde{n}}
P_{\beta_{\tilde{n}}}(U^{\dagger\,q-1})^{\tilde{n}}P_{\beta_{q\tilde{n}}}\right]\propto
\delta_{\alpha_{q\tilde{n}}\beta_{q\tilde{n}}}
\delta_{\alpha_{\tilde{n}}\beta_{\tilde{n}}}
\end{equation}
for all $q\in\natuerl$, {any 
$\rho_0\in\mathcal{S}_{\{P_{\mu}\}}$,} and
arbitrary $\alpha_{\tilde{n}}$, $\beta_{\tilde{n}}$, $\alpha_{q\tilde{n}}$, $\beta_{q\tilde{n}}$. 
In order to proceed we will need the 
following Lemma~\cite{Scherer2004A}.

\noindent 
{\bf Lemma:}
{\it Let} $\mathcal{H}$ {\it be a finite dimensional Hilbert space 
and}  $U:\cal H\rightarrow\cal H$ {\it any unitary map on} 
${\cal H}$. {\it Then} $\forall\,\epsilon >0\;\;\exists\,
q\in\natuerl$ {\it such that} $\parallel  U^q-\one_{\cal H} \parallel 
< \epsilon\,$, $\parallel \cdot \parallel$ {\it meaning the 
conventional operator norm, which 
is}  $\parallel\hspace{-1mm} A\hspace{-1mm} \parallel
:=\mbox{sup}\{\parallel\hspace{-1mm} Av \hspace{-1mm}\parallel\::\: 
v\in\mathcal{H}\,,\,\parallel\hspace{-1mm} v
\hspace{-1mm}\parallel=1\,\}$ {\it for an operator} $A$ {\it on} $\mathcal{H}$.

According to this Lemma, for any given arbitrarily small $\epsilon>0$ 
we can always find a $q\in\natuerl$ such that 
$U^q=\one_{\cal H}+\hat{\mathcal{O}}(\epsilon)$, where 
$\hat{\mathcal{O}}(\epsilon)$ is some operator whose 
norm is of order $\epsilon$: $\parallel 
\hat{\mathcal{O}}(\epsilon)\parallel<\epsilon$.
Using the submultiplicativity property 
of operator norms, we have 
\begin{equation}
\parallel U^{-1} \hat{\mathcal{O}}(\epsilon)\parallel\,\le 
\,\parallel U^{-1}\parallel\times
\parallel \hat{\mathcal{O}}(\epsilon)\parallel\,=\,
\parallel \hat{\mathcal{O}}(\epsilon)\parallel
\end{equation}
and hence $U^{q-1}=U^{-1}+\hat{\mathcal{O}'}(\epsilon)$, where
$\parallel\! \hat{\mathcal{O}'}(\epsilon)\!\parallel<\epsilon$.

Now we are in a position to derive a contradiction. 
We let our histories start with 
the initial state $\rho_0=\tilde{\rho}$.  
Furthermore we choose $\alpha_{\tilde{n}}=\mu'$,  
$\beta_{\tilde{n}}=\mu''$, and $\alpha_{q\tilde{n}}
=\beta_{q\tilde{n}}=\mu_0$.
Since $\mu'\ne\mu''$, condition~(\ref{nec_con_(q-1)}) becomes
\begin{equation} \label{beforelemma}
\forall q\in\natuerl\,:\; 
\mbox{Tr}\left[P_{\mu_0}(U^{q-1})^{\tilde{n}}P_{\mu'}U^{\tilde{n}} 
\,\tilde{\rho}\, U^{\dagger\,{\tilde{n}}}
P_{\mu''}(U^{\dagger\,q-1})^{\tilde{n}}P_{\mu_0}\right] = 0 \;.
\end{equation}
Choosing $q$ such that $\parallel  U^q-\one_{\cal H} \parallel 
< \epsilon\,$ for a given {\em arbitrarily small} $\epsilon> 0$, 
we get a situation where the expressions $(U^{q-1})^{\tilde{n}}$ 
and $(U^{\dagger\,q-1})^{\tilde{n}}$ in Eq.~(\ref{beforelemma}) 
can be replaced by
$(U^{\dagger}+\hat{\mathcal{O}'}(\epsilon))^{\tilde{n}}$ 
and $(U+\hat{\mathcal{O}'}^{\dagger}(\epsilon))^{\tilde{n}}$, 
respectively. In the following it will be convenient to use the 
definition
\begin{equation}
A_{r_1,r_2,\dots,r_{\tilde{n}}}:=\prod_{i=1}^{\tilde{n}}\left(
U^{\dagger\,r_i}(\hat{\mathcal{O}'}(\epsilon))^{1-r_i}\right)\;,
\end{equation}
where the operators inside the product are written out from left 
to right in the order of increasing index $i$. Using this definition 
we have:
\begin{equation} 
(U^{\dagger}+\hat{\mathcal{O}'}(\epsilon))^{\tilde{n}}
=\sum_{r_1,\dots,r_{\tilde{n}}\in\{0,1\}}\hspace{-2mm}
A_{r_1,\dots,r_{\tilde{n}}}\;.
\end{equation}
This yields for the left hand side of Eq.~(\ref{beforelemma}): 
\begin{equation*}\hspace{-6cm}
\mbox{Tr}\left[P_{\mu_0}(U^{q-1})^{\tilde{n}}P_{\mu'}U^{\tilde{n}} 
\,\tilde{\rho}\, U^{\dagger\,{\tilde{n}}}
P_{\mu''}(U^{\dagger\,q-1})^{\tilde{n}}P_{\mu_0}\right]
\end{equation*}
\begin{eqnarray} 
&=& 
\mbox{Tr}\Big[P_{\mu_0}\Big(\sum_{r_1,\dots,r_{\tilde{n}}\in\{0,1\}}\hspace{-2mm}
A_{r_1,\dots,r_{\tilde{n}}}\Big)P_{\mu'}U^{\tilde{n}}\,\tilde{\rho}\:
U^{\dagger\,{\tilde{n}}}P_{\mu''}\Big(\sum_{s_1,\dots,
s_{\tilde{n}}\in\{0,1\}}\hspace{-2mm}A^{\dagger}_{s_1,\dots,s_{\tilde{n}}}\Big)
P_{\mu_0}\Big]\nonumber\\
&=&\sum_{r_1,\dots,r_{\tilde{n}}\in\{0,1\}}\:
\sum_{s_1,\dots,s_{\tilde{n}}\in\{0,1\}}
\mbox{Tr}\Big[P_{\mu_0}A_{r_1,\dots,r_{\tilde{n}}}
P_{\mu'}U^{\tilde{n}}\,\tilde{\rho}\: U^{\dagger\,{\tilde{n}}}
P_{\mu''}A^{\dagger}_{s_1,\dots,s_{\tilde{n}}}
P_{\mu_0}\Big]\;.
\end{eqnarray}

\noindent
According to (\ref{beforelemma}) the left hand side of this equation
must be zero. Hence we have:
\begin{equation*}\hspace{-6cm}
\mbox{Tr}\left[P_{\mu_0}(U^{\dagger})^{\tilde{n}}
P_{\mu'}U^{\tilde{n}}\,\tilde{\rho}\: U^{\dagger\,{\tilde{n}}}
P_{\mu''}U^{\tilde{n}}P_{\mu_0}\right]
\end{equation*}
\begin{eqnarray} 
&=& 
-\sum_{r_1,\dots,r_{\tilde{n}}\in\{0,1\}\atop
r_1+\dots+r_{\tilde{n}}<\tilde{n}}\:
\sum_{s_1,\dots,s_{\tilde{n}}\in\{0,1\}
\atop{s_1+\dots+s_{\tilde{n}}<\tilde{n}\atop\,}}
\mbox{Tr}\Big[P_{\mu_0}A_{r_1,\dots,r_{\tilde{n}}}
P_{\mu'}U^{\tilde{n}}\,\tilde{\rho}\: U^{\dagger\,{\tilde{n}}}
P_{\mu''}A^{\dagger}_{s_1,\dots,s_{\tilde{n}}}
P_{\mu_0}\Big]\;.
\end{eqnarray}

\noindent
Using the cyclic permutation-invariance property of the trace 
and the triangle inequality, we obtain
\begin{eqnarray} \label{eq:triangle}
&&\left|\mbox{Tr}\left[P_{\mu''}(U^{\tilde{n}}
P_{\mu_0}U^{\dagger\,\tilde{n}})P_{\mu'}
(U^{\tilde{n}}\,\tilde{\rho}\,U^{\dagger\,{\tilde{n}}})
P_{\mu''}\right]\right|\le\nonumber\\
&&\le
\sum_{r_1,\dots,r_{\tilde{n}}\in\{0,1\}\atop
r_1+\dots+r_{\tilde{n}}<\tilde{n}}\:
\sum_{s_1,\dots,s_{\tilde{n}}\in\{0,1\}
\atop{s_1+\dots+s_{\tilde{n}}<\tilde{n}\atop\,}}
\Big|\mbox{Tr}\Big[A^{\dagger}_{s_1,\dots,s_{\tilde{n}}}P_{\mu_0}
A_{r_1,\dots,r_{\tilde{n}}}
P_{\mu'}U^{\tilde{n}}\,\tilde{\rho}\: U^{\dagger\,{\tilde{n}}}
P_{\mu''}\Big]\Big|\;.
\end{eqnarray}
Utilizing the inequality
$\:\left|\,\mbox{Tr}[BT]\,\right|\le\,\parallel B \parallel
\mbox{Tr}\sqrt{T^{\dagger}T}\:$
for bounded operators $B:\cal H\rightarrow\cal H$ and 
operators $T:\cal H\rightarrow\cal H$ with finite trace 
norm $\parallel T \parallel_1:=\mbox{Tr}\sqrt{T^{\dagger}T}$, see Ref.~\cite{Alicki2001}, 
we deduce from Eq.~(\ref{eq:triangle}):
\begin{eqnarray} \label{eq:norm-tracenorm-ineq}
&&\left|\mbox{Tr}\left[P_{\mu''}(U^{\tilde{n}}
P_{\mu_0}U^{\dagger\,\tilde{n}})P_{\mu'}
(U^{\tilde{n}}\,\tilde{\rho}\,U^{\dagger\,{\tilde{n}}})
P_{\mu''}\right]\right|\le\nonumber\\
&&\le
\sum_{r_1,\dots,r_{\tilde{n}}\in\{0,1\}\atop
r_1+\dots+r_{\tilde{n}}<\tilde{n}}\:
\sum_{s_1,\dots,s_{\tilde{n}}\in\{0,1\}
\atop{s_1+\dots+s_{\tilde{n}}<\tilde{n}\atop\,}}
\parallel B^{s_1,\dots,s_{\tilde{n}}}_{r_1,\dots,r_{\tilde{n}}}\parallel
\mbox{Tr}\sqrt{T^{\dagger}T}\;,
\end{eqnarray}
where we defined 
\begin{eqnarray}
B^{s_1,\dots,s_{\tilde{n}}}_{r_1,\dots,r_{\tilde{n}}}&:=&
A^{\dagger}_{s_1,\dots,s_{\tilde{n}}}P_{\mu_0}
A_{r_1,\dots,r_{\tilde{n}}}\;,\\
T&:=&P_{\mu'}U^{\tilde{n}}\,\tilde{\rho}\: U^{\dagger\,{\tilde{n}}}
P_{\mu''}\;.
\end{eqnarray}
Using the fact that $\parallel B^{\dagger} \parallel =\parallel B
\parallel $ for any bounded operator $B$ and it's adjoint
$B^{\dagger}$ \cite{Weidmann1980}, we have $\parallel \hat{\mathcal{O}'}
^{\dagger}(\epsilon) \parallel =
\parallel \hat{\mathcal{O}'}(\epsilon) \parallel<\epsilon$.
Utilizing the submultiplicativity property 
of operator norms we deduce that the norms of the operators
$B^{s_1,\dots,s_{\tilde{n}}}_{r_1,\dots,r_{\tilde{n}}}$ are all
bounded from above by $\epsilon$, except in the case where 
all $s_1,\dots,s_{\tilde{n}}$ and all $r_1,\dots,r_{\tilde{n}}$ 
are equal $1$, which is excluded from the sum on the right-hand 
side of Eq.~(\ref{eq:norm-tracenorm-ineq}). Indeed we have:
\begin{eqnarray}
\parallel B^{s_1,\dots,s_{\tilde{n}}}_{r_1,\dots,r_{\tilde{n}}}\parallel
&\le&
\left(\prod_{i=1}^{\tilde{n}}\parallel U\parallel^{s_i}\,
\parallel \hat{\mathcal{O}'}^{\dagger}(\epsilon))\parallel^{1-s_i}\right)
\parallel P_{\mu_0} \parallel 
\left(\prod_{i=1}^{\tilde{n}}\parallel U^{\dagger}\parallel^{r_i}\,
\parallel \hat{\mathcal{O}'}(\epsilon)) \parallel^{1-r_i}\right)
\nonumber\\
&\le&\left(\prod_{i=1}^{\tilde{n}}\epsilon^{1-s_i}\right)
\left(\prod_{j=1}^{\tilde{n}}\epsilon^{1-r_j}\right)\nonumber\\
&\le&\epsilon^2<\epsilon\;,\quad\mbox{if}\quad 
s_1+\dots+s_{\tilde{n}}<\tilde{n}\:,\: r_1+\dots+r_{\tilde{n}}<\tilde{n} \;,
\end{eqnarray}
where we used  
$\parallel P_{\mu_0}\parallel=\parallel U\parallel=\parallel
U^{\dagger}\parallel=1$  and $\epsilon\ll 1$. 
With the definition $M:=\mbox{Tr}\sqrt{T^{\dagger}T}$ 
we finally conclude from Eq.~(\ref{eq:norm-tracenorm-ineq}):
\begin{equation}
\left|\mbox{Tr}\left[P_{\mu''}(U^{\tilde{n}}
P_{\mu_0}U^{\dagger\,\tilde{n}})P_{\mu'}
(U^{\tilde{n}}\,\tilde{\rho}\,U^{\dagger\,{\tilde{n}}})
P_{\mu''}\right]\right|
<2^{2{\tilde{n}}}M\epsilon\:.
\end{equation}
Since $c$, $\tilde{n}$ and $M$ are fixed  
constants, we can always arrange $2^{2{\tilde{n}}}M\epsilon
<\left|c\right|$ by choosing a sufficiently small $\epsilon>0$. 
This contradicts the assumption~(\ref{eq:Tr[...]=c}) 
and thus proves our proposition~(\ref{eq:loopcondition}).

We are now in a position to derive the commutativity 
condition (b) of the theorem. It is a straightforward 
consequence of proposition (\ref{eq:loopcondition}) 
we have just proven. Taking condition (\ref{eq:loopcondition}) 
and choosing in it the state $\rho\in\mathcal{S}_{\{P_{\mu}\}}$ 
to be proportional to the projector sandwiched between $U^n$ and 
$U^{\dagger\,n}$ within the first bracket, 
\begin{equation} 
\rho=\frac{P_{\mu_0}}{\mbox{Tr}\left[P_{\mu_0}\right]}\;,
\end{equation}
where $P_{\mu_0}$ is still arbitrary, we necessarily get the condition
\begin{eqnarray} \label{eq:specialloopcondition}
&&
\forall\, n\in\natuerl\:\:
\forall\,\mu_0,\mu',\mu''\quad\mbox{with}\quad
\mu'\not=\mu'':\nonumber\\
&&
\mbox{Tr}\left[P_{\mu''}(U^nP_{\mu_0}U^{\dagger\,n})P_{\mu'}
(U^nP_{\mu_0} U^{\dagger\,n})P_{\mu''}\right]=0\:.
\end{eqnarray}
With the definition $A:=P_{\mu'}
(U^nP_{\mu_0} U^{\dagger\,n})P_{\mu''}$ 
Eq.~(\ref{eq:specialloopcondition}) becomes 
$\mbox{Tr}\left[A^{\dagger}A\right]=0$. 
Since $A^{\dagger}A$ is a positive operator, this is possible if and only 
if $A=0$. Hence condition (\ref{eq:specialloopcondition}) is equivalent
to 
\begin{eqnarray} \label{eq:specialloopcondition2}
&&\forall\, n\in\natuerl\:\:
\forall\,\mu_0,\mu',\mu''\quad\mbox{with}\quad
\mu'\not=\mu'':\nonumber\\
&&
P_{\mu'}
(U^nP_{\mu_0} U^{\dagger\,n})P_{\mu''}=0\;.
\end{eqnarray}
This condition implies   
\begin{equation} 
\sum_{\mu'}P_{\mu'}(U^nP_{\mu_0} U^{\dagger\,n})P_{\mu''}=
P_{\mu''}(U^nP_{\mu_0} U^{\dagger\,n})P_{\mu''}
\end{equation}
for any $\mu_0$ and $\mu''$, and arbitrary $n\in\natuerl$. 
But since $\sum_{\mu'}P_{\mu'}=\one_{\cal H}$, 
the left hand side of the last equation must be equal to $(UP_{\mu_0} U^{\dagger})P_{\mu''}$. 
Hence we obtain 
\begin{equation} \label{eq:P(UPU^)P=(UPU^)P}
P_{\mu''}(U^nP_{\mu_0} U^{\dagger\,n})P_{\mu''}
=(U^nP_{\mu_0} U^{\dagger\,n})P_{\mu''}
\end{equation}
on the one hand and by taking the adjoint of Eq.~(\ref{eq:P(UPU^)P=(UPU^)P})
\begin{equation} 
P_{\mu''}(U^nP_{\mu_0} U^{\dagger\,n})P_{\mu''}
=P_{\mu''}(U^nP_{\mu_0} U^{\dagger\,n})
\end{equation}
on the other hand, for any $n\in\natuerl$ and arbitrary $\mu_0$ and $\mu''$. 
Therefore 
\begin{equation} 
(U^nP_{\mu_0} U^{\dagger\,n})P_{\mu''}=P_{\mu''}(U^nP_{\mu_0} U^{\dagger\,n})
\end{equation}
for any $n\in\natuerl$ and 
arbitrary $\mu_0$, $\mu''$, and so 
$\left[U^nP_{\mu_0} U^{\dagger\,n}\,,\,P_{\mu''}\right]=0$ 
for any $n\in\natuerl$ and all $P_{\mu_0}, P_{\mu''}\in\{P_{\mu}\}$. $\Box$

The implication (a)$\Rightarrow $(c) of the theorem constitutes the 
main result of this paper: the decoherence of histories of arbitrary
length for all initial states from the set 
${\cal S}_{\{P_\mu\}}$ implies decoherence  
of such histories for arbitrary initial states 
$\rho\in\mathcal{S}$. 
It should be mentioned that the set ${\cal S}_{\{P_\mu\}}$
can be viewed as the smallest natural set of states that 
is associated with our framework. It is discrete and
may consist of just two elements (in the case of \lq\lq yes-no'' propositions).
The set ${\cal S}$, on the other hand, contains the continuum
of all possible states that are allowed in our framework.

In \cite{Scherer2004A} the notion 
of classical states with respect to  a partition 
$\{P_{\mu}\}$ was introduced:\newline
\noindent {\bf Definition 4:}
A state represented by the density operator $\rho$ is called 
{\em classical with respect to (w.r.t.) a partition  $\{P_{\mu}\}$}
of the Hilbert space $\cal H$, if it is block-diagonal
w.r.t.\ $\{P_\mu\}$, i.e., if 
$\rho = \sum_{\mu}P_{\mu}\,\rho\, P_{\mu}\:$. We denote by 
$\mathcal{S}^{\mbox{\scriptsize cl}}_{\{P_{\mu}\}}$
the set of all density operators that are classical
w.r.t.~$\{P_{\mu}\}$. 

In~\cite{Scherer2004A} it was shown that in the case of 
fine-grained partitions sets of histories of arbitrary length decohere 
for all classical initial states {\em only if} the unitary dynamics 
preserves the classicality of states, i.e.\ {\em only if} 
\begin{equation} \label{ClassicalityPreservation} 
\forall\rho\in\mathcal{S}^{\mbox{\scriptsize cl}}_{\{P_{\mu}\}}\,:\;
U\rho U^{\dagger}\in \mathcal{S}^{\mbox{\scriptsize
    cl}}_{\{P_{\mu}\}}.
\end{equation} 
It is a single-iteration criterion: to verify that it  
holds for a particular unitary map $U$, only a single iteration of the map 
has to be taken into account, which can be much easier than establishing 
decoherence directly by computing the off-diagonal elements of the decoherence
functional. This is especially useful for studying chaotic quantum maps, 
for which typically only the first iteration is known in a closed 
analytical form~\cite{Soklakov2000a}. 
Unfortunately, condition~(\ref{ClassicalityPreservation}) fails to
be necessary in the coarse-grained case. The following simple
corollary of our theorem provides a necessary single-iteration condition 
that applies to arbitrary coarse-grainings and is equivalent 
to~(\ref{ClassicalityPreservation}) in the fine-grained case.

\noindent{\bf Corollary:}
{\it Let a projective partition}  $\{P_{\mu}\}$ 
{\it of a finite dimensional Hilbert space $\cal H$ and a unitary 
map} $U$ {\it on} $\cal H$ {\it be given. The medium decoherence condition is
then satisfied for all classical initial states and arbitrarily long
histories, i.e.,}
\begin{equation}  
\forall \,\rho\in\mathcal{S}^{\mbox{\scriptsize cl}}_{\{P_{\mu}\}}\;
\forall\, k\in\hspace{-0.5mm}\natuerl \;\;\forall\,h_{\balphas},
h_{\bbetas}\in\mathcal{K}[\{P_{\mu}\}\,;\,k\,]:
\;\mathcal{D}_{U,\rho}\,[h_{\balphas},h_{\bbetas}]
\propto  \delta_{\balphas\bbetas}\;,
\end{equation}
{\em only if the following necessary condition is fulfilled:} 
\begin{equation}  
\forall\,P_{\mu'}, P_{\mu''}\in\{P_{\mu}\}
\,:\;\;\left[UP_{\mu'}U^{\dagger}\,,\,P_{\mu''}\right]=0\,.
\label{nec_singleiteration_comm_cond}
\end{equation}\\
\noindent{\bf Proof:}
follows trivially from the implication (a)$\Rightarrow$(b)  
of the theorem, as $\mathcal{S}_{\{P_{\mu}\}}\subset
\mathcal{S}^{\mbox{\scriptsize cl}}_{\{P_{\mu}\}}$.  $\Box$\\

In summary, we investigated decoherence properties of sets of quantum
histories constructed from a fixed projective partition $\{P_\mu\}$ of a
finite dimensional Hilbert space. We found that if decoherence
is established for arbitrary history lengths and all
initial states from ${\cal S}_{\{P_\mu\}}$, which is
the smallest natural set induced by $\{P_{\mu}\}$,  
then any set of histories constructed from $\{P_\mu\}$
is decoherent for all possible initial states.
In addition, we provided a necessary single-iteration 
criterion for decoherence of arbitrarily long
histories that generalizes the condition
of~\cite{Scherer2004A} to the case of arbitrary coarse-grainings.

\noindent{\bf Acknowledgements:} We would like to thank R\"udiger Schack 
for helpful discussions.

\end{document}